\documentclass[superscriptaddress, reprint]{revtex4-1}
\usepackage{hyperref}
\usepackage[version=4]{mhchem} 
\usepackage{graphicx}
\usepackage{chemmacros}
\usepackage{gensymb}
\usepackage{dcolumn}
\usepackage{amssymb}
\begin{document}

\title{The magnetic structure and spin-flop transition in the A-site columnar-ordered quadruple perovskite \ce{TmMn3O6}}

\author{A. M. Vibhakar}
\affiliation{Clarendon Laboratory, Department of Physics, University of Oxford, Oxford, OX1 3PU, United Kingdom}
\author{D. D. Khalyavin}
\affiliation{ ISIS facility, Rutherford Appleton Laboratory-STFC, Chilton, Didcot, OX11 0QX, United Kingdom}
\author{P. Manuel}
\affiliation{ ISIS facility, Rutherford Appleton Laboratory-STFC, Chilton, Didcot, OX11 0QX, United Kingdom}
\author{L. Zhang}
\affiliation{Graduate School of Chemical Sciences and Engineering, Hokkaido University, North 10 West 8, Kita-ku, Sapporo, Hokkaido 060-0810, Japan}
\affiliation{Research Center for Functional Materials, National Institute for Materials Science (NIMS), Namiki 1-1, Tsukuba, Ibaraki 305-0044, Japan}
\author{K. Yamaura}
\affiliation{Graduate School of Chemical Sciences and Engineering, Hokkaido University, North 10 West 8, Kita-ku, Sapporo, Hokkaido 060-0810, Japan}
\affiliation{Research Center for Functional Materials, National Institute for Materials Science (NIMS), Namiki 1-1, Tsukuba, Ibaraki 305-0044, Japan}
\author{P. G. Radaelli}
\affiliation{Clarendon Laboratory, Department of Physics, University of Oxford, Oxford, OX1 3PU, United Kingdom}
\author{A. A. Belik}
\affiliation{Research Center for Functional Materials, National Institute for Materials Science (NIMS), Namiki 1-1, Tsukuba, Ibaraki 305-0044, Japan}
\author{R. D. Johnson}
\email{roger.johnson@physics.ox.ac.uk}
\affiliation{Clarendon Laboratory, Department of Physics, University of Oxford, Oxford, OX1 3PU, United Kingdom}

\date{\today}

\begin{abstract}
We present the magnetic structure of \ce{TmMn3O6}, solved via neutron powder diffraction --- the first such study of any \ce{$R$Mn3O6} A-site columnar-ordered quadruple perovskite to be reported. We demonstrate that long range magnetic order develops below 74 K, and at 28 K a spin-flop transition occurs driven by $f$-$d$ exchange and rare earth single ion anisotropy. In both magnetic phases the magnetic structure may be described as a collinear ferrimagnet, contrary to conventional theories of magnetic order in the manganite perovskites. Instead, we show that these magnetic structures can be understood to arise due to ferro-orbital order, the A, A$'$ and A$''$ site point symmetry,  $mm2$, and the dominance of A-B exchange over both A-A and B-B exchange, which together are unique to the \ce{$R$Mn3O6} perovskites.
\end{abstract}
 
\maketitle

\section{Introduction}
The ordering of charges, orbitals, and spins in the perovskite manganites (general chemical formula \ce{ABO3}, B = Mn) gives rise to a multitude of different electronic phases \cite{1999Coey}. For example, the \ce{$R$MnO3} simple perovskite manganites ($R$ = rare earth or yttrium) and their derivatives are known to support ferromagnetism \cite{1950Jonker, 1956Jonker}, collinear antiferromagnetism \cite{1955Wollan}, and complex non-collinear magnetic structures \cite{1996Chen}, which can couple to charge and structural degrees of freedom giving rise to technologically important materials properties such as metal-to-insulator transitions \cite{1984Ganguly}, multiferroicity \cite{2003Kimura} and colossal magnetoresistance \cite{1994McCormack}. The variety of magnetic structures found in the simple perovskite manganites are primarily established via magnetic interactions between the $d$ electrons of B site manganese ions (B-B exchange), and understood in terms of the Goodenough-Kanamori-Anderson rules \cite{1963Goodenough}. In addition, low temperature magnetic phase transitions, such as spin reorientation transitions \cite{1964Koehler, 1999Bieringer, 2001Munoz}, can occur due to lower energy $f$-$d$ (A-B exchange) and $f$-$f$ (A-A exchange) interactions if a magnetic rare earth ion is present.

Equally, the \ce{AMn7O12} quadruple perovskite family (general chemical formula \ce{AA$'$3B4O12}, A$'$ = B = Mn) support charge, orbital, and spin long range order \cite{2004Prodi,2007Vasil,2014Gilioli} and host properties of low field magnetoresistance \cite{1999Zeng} and multiferroicity \cite{2009Mezzadri, 2012Johnson}. However, compared to the simple perovskites the introduction of manganese onto the A site sublattice via a pattern of large $a^+a^+a^+$ octahedral tilts (in Glazer notation \cite{1972GlazerNotation}) has a profound effect on the magnetic exchange interactions. Firstly, the average A-A and A-B exchange interactions can be greatly enhanced due to the inclusion of $3d$ transition metal ions onto the perovskite A sites, as well as creating new exchange pathways for geometric frustration. Secondly, the B-B exchange interaction can be diminished as a result of the severe octahedral tilting \cite{2018Johnson_1}. It has been shown that such complexity leads to members of this family adopting rather unique magnetic ground states, for example, where an incommensurate magneto-orbital coupling gives rise to a constant moment magnetic helix but with a modulated spin helicity \cite{2016Johnson, 2017Johnson}, classic \emph{pseudo}-CE-type magnetic structures that develop an incommensurate spin rotation \cite{2018Johnson_2}, and commensurate ferrimagnetism that supplants conventional B site antiferromagnetism \cite{2018Johnson_1}.

The newly synthesised \ce{$R$Mn3O6} series ($R$ = Gd-Tm and Y have been synthesised to date \cite{2017Zhang}) belong to the wider family of triple A-site columnar-ordered quadruple perovskites of stoichiometry \ce{A2A$'$A$''$B4O12}. Here, the triple order of A site cations affords a new paradigm in the study of magnetism and structure-property relationships in the perovskite manganites. As in the \ce{AMn7O12} quadruple perovskites, B-B exchange is likely to be suppressed due to large octahedral tilts, but so too are A-A exchange interactions owing to a fundamentally different bonding geometry. Hence, novel magnetic ground states are expected to arise, primarily established by A-B exchange interactions, and selectively tuned by choice of A, A$'$ and A$''$ cations (N.B. \textquoteleft A-A{\textquoteright} and {\textquoteleft}A-B{\textquoteright} denote average exchange interactions involving all A, A$'$ and A$''$ ions). 

\begin{figure}
\includegraphics[width=\linewidth]{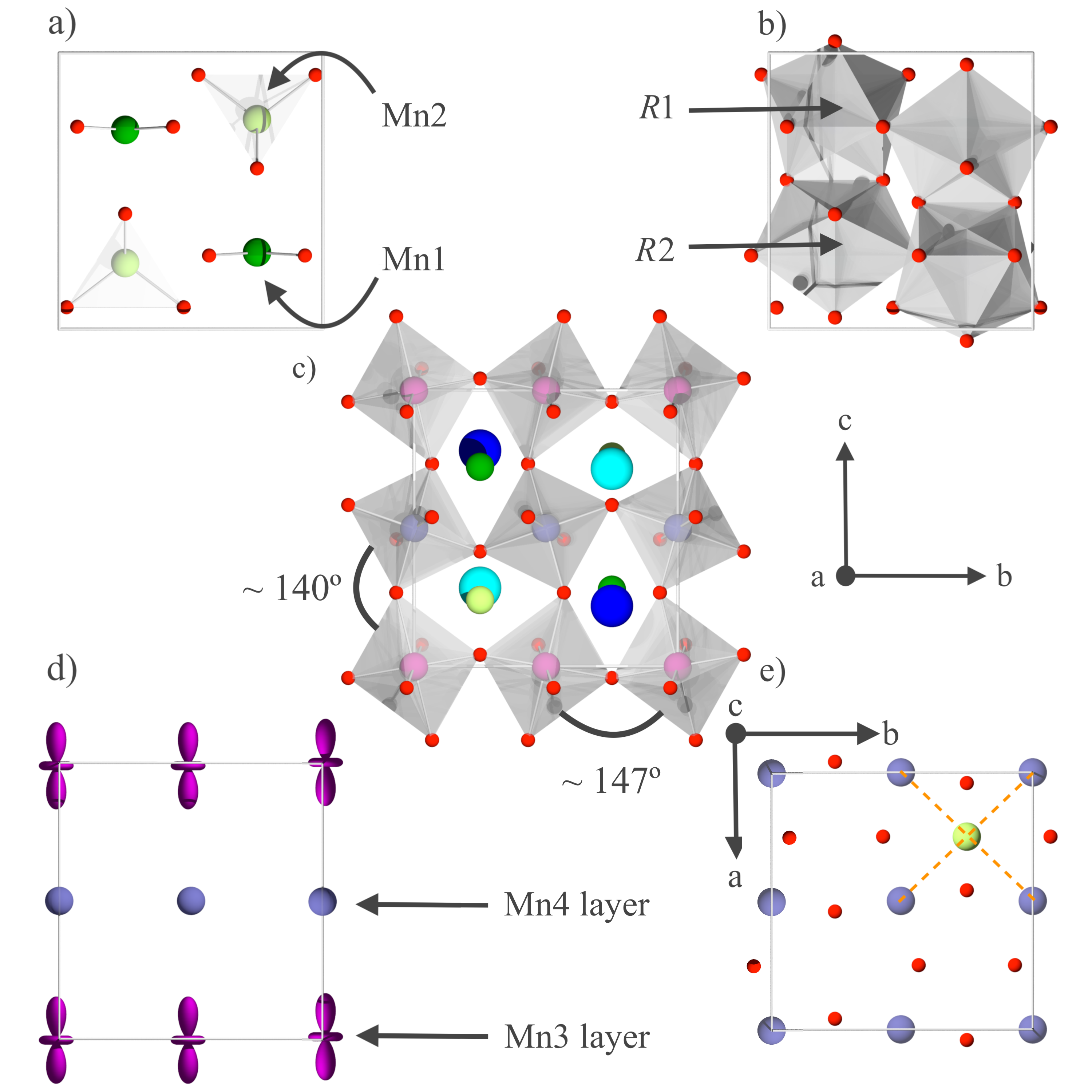}
\caption{The crystal structure of \ce{$R$Mn3O6} in the $Pmmn$ space group a) - d) a projection of the crystal structure as seen from the $a$ axis. a) The Mn1 (A$'$) and Mn2 (A$''$) sites in their square planar and tetrahedral oxygen coordinations, respectively. b) The 10-fold coordinated {\it{R}} sites. c) The unit cell with \ce{MnO6} octahedra drawn to illustrate the $a^+a^+c^-$ tilting pattern. d) The Mn3 and Mn4 ions (B sites) with the occupied $d_{3z^2-r^2}$ orbitals drawn in purple. e) A section of the crystal structure as viewed from the c-axis showing a plane of Mn4 ions and a single Mn2 site. The orange dashed lines represent the A-B exchange interactions which are equivalent by the $mm2$ symmetry of the Mn2 site.}
\label{FIG::crystruc}
\end{figure}

The crystal structure of \ce{$R$Mn3O6}  has orthorhombic symmetry with space group $Pmmn$ and a large $a^+a^+c^-$ tilting pattern (in Glazer notation \cite{1972GlazerNotation}), which gives B-O-B bond angles between $\sim$135 and 150{\degree} \citep{2017Zhang}, and accommodates the triple A site order by which half (A) are occupied by $R^{3+}$, a quarter (A$'$) by Mn$^{3+}$, labeled Mn1, and the remaining quarter (A$''$) by Mn$^{2+}$, labeled Mn2. $R$ ions occupy two 10-fold coordinated, symmetry inequivalent sites labeled $R$1 and $R$2 (shown in Figure \ref{FIG::crystruc}b), which form columns along $c$. The Mn A$'$ and A$''$ sites form $c$ axis chains alternating between square planar and tetrahedral oxygen coordinations, as shown in Figure \ref{FIG::crystruc}a. This difference in the coordination environment of the Mn1 and Mn2 ions requires them to have +3 and +2 oxidations states, respectively. Furthermore, the formation of alternating square planar and tetrahedral coordinations removes nearest-neighbour magnetic super exchange pathways between manganese A$'$ and A$''$ sites, thus suppressing A-A exchange interactions. The B sites are equally occupied by alternating layers of Mn in the +3 and +3.5 oxidation states \citep{2017Zhang}, which we label Mn3 and Mn4, respectively. The isocharge layers are stacked along the $c$ axis, as shown in Figure \ref{FIG::crystruc}d, which gives rise to an unusual form of ferro-orbital order, where the half occupied, Mn3 $d_{3z^2-r^2}$ orbitals are all aligned approximately parallel to $c$ \cite{2018BelikTriple}, mediated by the cooperative Jahn-Teller effect.

In this paper, we show that below $T_\mathrm{c} = 74$ K \ce{TmMn3O6} develops long range ferrimagnetic order of the A$'$, A$''$ and B site Mn spins. On further cooling, the system undergoes a spin flop transition at $T_\mathrm{flop} = 28$ K where all Mn spins spontaneously rotate by 90$^\circ$ from parallel to the $c$ axis, to parallel to the $b$ axis, concomitant with the development of a polarised moment on the Tm ions --- similar to the low temperature spin reorientation transition observed in the columnar-ordered double perovskite \ce{MnNdMnSbO6} \cite{2018Solana}. Through model spin Hamiltonian calculations of the Tm sublattice that include a point-charge approximation of the Tm crystal electric field, we show that the spin reorientation transition is driven by $f$-$d$ exchange and an Ising-like single-ion anisotropy of Tm. Remarkably, in both magnetically ordered phases of \ce{TmMn3O6}, the ferrimagnetic structure is in contradiction with theoretical predictions based on the Goodenough-Kanamori-Anderson rules. Instead, we show that these magnetic structures can be understood to arise due to the ferro-orbital order, A, A$'$ and A$''$ site symmetry, and the dominance of A-B exchange over both A-A and B-B exchange, which taken together are unique to the \ce{$R$Mn3O6} perovskites. The paper is organised as follows. We present the experimental details in Section \ref{SEC::experimentaldetails}, the results of neutron powder diffraction data analysis, crystal electric field, and single ion anisotropy calculations in Section \ref{SEC::results}, a discussion on the formation and thermal evolution of the \ce{TmMn3O6} magnetic structures in Section \ref{SEC::discussion}, and we summarise our conclusions in Section \ref{SEC::conclusions}.

\section{Experimental Details}\label{SEC::experimentaldetails}
Polycrystalline samples with nominal composition \ce{Tm_{0.91}Mn_{3.09}MnO6}  (hereafter referred to as \ce{TmMn3O6}) were prepared from stoichiometric mixtures of \ce{Mn2O3} and \ce{Tm2O3} (99.9\%). Single-phase \ce{Mn2O3} was prepared from commercial \ce{MnO2} (99.99\%) by heating in air at 923 K for 24 h. The mixtures were placed in Pt capsules and treated at 6 GPa and $\sim$1670 K for 2 h (the time required to heat the sample to the desired temperature was 10 min) in a belt-type high-pressure apparatus. After the heat treatments, the samples were quenched to room temperature, the pressure was slowly released, and a black pellet of loosely packed powder was obtained. The temperature of the high-pressure apparatus was controlled by the heating power with a calibrated relationship between power and temperature. We emphasize that a sample with nominal composition \ce{TmMn3O6} contained a large amount of \ce{Tm_{1-x}Mn_{x}MnO3} impurity (approximately 28 wt. \%). Shifting the nominal composition to  \ce{Tm_{0.91}Mn_{3.09}MnO6} was found to significantly reduce the impurity content.

DC magnetisation measurements were performed using a SQUID magnetometer (Quantum Design, MPMS-XL-7T) between 2 and 400 K in different applied fields under both zero-field-cooled (ZFC) and field-cooled on cooling (FCC) conditions.

Neutron diffraction measurements were performed using WISH \citep{2011ChaponWISH}, a time of flight diffractometer at ISIS. A 1.5 g powder sample of \ce{TmMn3O6} was loaded into a 6 mm diameter vanadium can, and mounted within a $^4$He cryostat. The sample was cooled to 1.5 K and diffraction data with good counting statistics were collected upon warming at 5 K intervals, except between 70 and 80 K where measurements were taken every 2 K. Data were also collected with high counting statistics within each magnetic phase including the paramagnetic phase for reference (85 K, 40 K and 1.5 K). Symmetry analysis was performed using the \textsc{isotropy} software suite \citep{2006Isodistort}, and the diffraction data were fit using \textsc{fullprof} \cite{1993Rodriguez}. 

\section{Results}\label{SEC::results}
\subsection{Crystal and Magnetic Structures}

\begin{figure}
\includegraphics[width=0.9\linewidth]{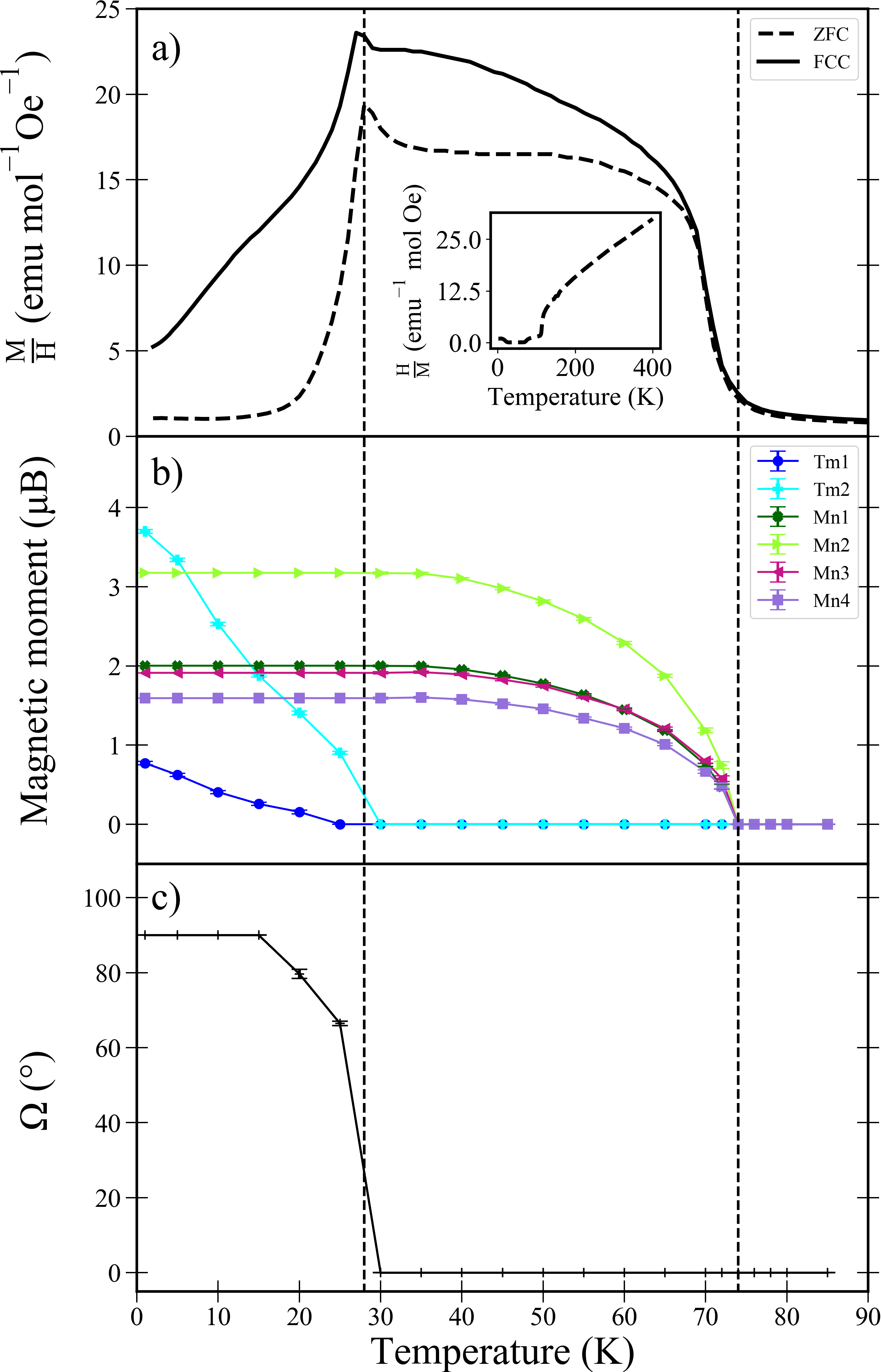}
\caption{Temperature dependence of a) ZFC and FCC magnetisation measurements under an applied DC field of 100 Oe, b) single ion moment magnitudes given for each of the magnetic sublattices, and c) the direction of magnetisation, where $\Omega$ is the angle that describes a rotation from the $c$ axis to the $b$ axis. The black dashed lines are drawn at $T_\mathrm{c}$ =  74 K,  and at $T_\mathrm{flop}$ = 28 K. The inset in a) shows the temperature dependence of $H/M$ up to 400K.}
\label{FIG::tempdep}
\end{figure}

Variable temperature ZFC and FCC DC magnetisation measurements are shown in Figure \ref{FIG::tempdep}a, which are consistent with previously reported AC magnetic susceptibility data \cite{2017Zhang}. Two magnetic phase transitions were observed, one at 74 K, below which a rapid increase in the susceptibility indicates the presence of ferromagnetic sublattices, and a second at 28 K, marked by a cusp-like increase in the susceptibility, followed by a gradual decrease that persists down to the lowest temperatures measured. This behaviour can be fully understood in light of the magnetic structures determined by neutron powder diffraction, and will be discussed at the end of this section.

\begin{figure}
\includegraphics[width=\linewidth]{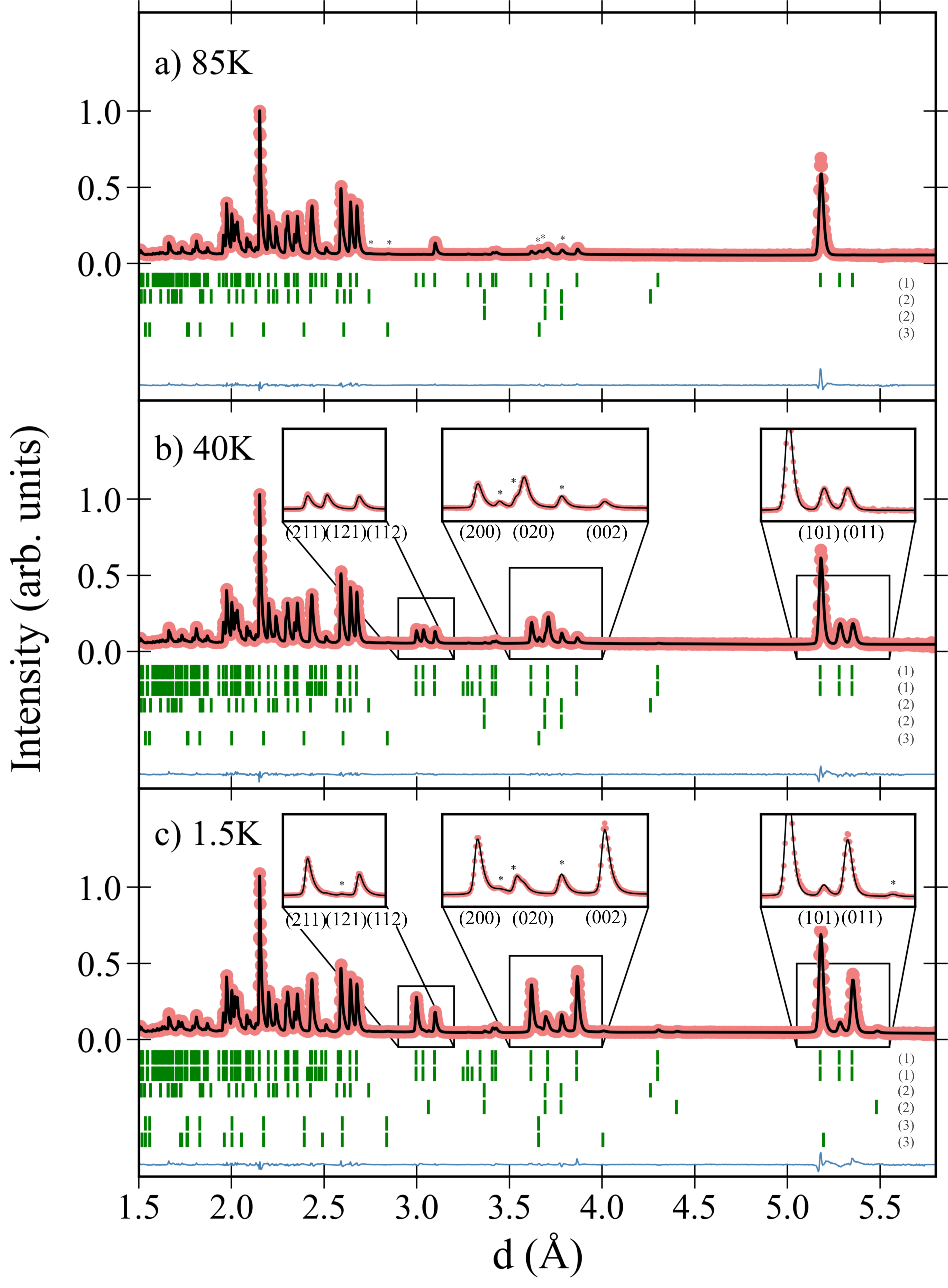}
\caption{Neutron powder diffraction data taken on WISH at temperatures of a) 85 K b) 40 K and c) 1.5 K. The refinement is given by the black line and the raw data by the red circles, peaks labeled with an asterisk are from the impurity phases. The green tick marks indicate, from top to bottom, the nuclear and magnetic reflections from  \ce{TmMn3O6}, labeled (1), the \ce{TmMnO3}-related impurity, labeled (2), and the \ce{MnCO3} impurity, labeled (3).}
\label{FIG::neutrondata}
\end{figure}

A $Pmmn$ crystal structure model based upon the published structure of \ce{DyMn3O6} \cite{2017Zhang} was refined against neutron powder diffraction data measured in the paramagnetic phase at 85K (Figure \ref{FIG::neutrondata}a). The occupation of all atomic sites was set to one, but Mn was free to substitute Tm --- as expected for the \ce{$R$Mn3O6} phases \cite{2018Belik}. An excellent fit was achieved ($R = 3.32\%$, $wR = 2.68\%$, $R_\mathrm{Bragg} = 4.57\%$), and the refined crystal structure parameters are given in Table \ref{TAB::xyz_bvs}. Also tabulated are the cation bond valence sums, calculated using bond lengths given in Tables \ref{TAB::Asitebonds} and \ref{TAB::Bsitebonds} of Appendix \ref{SEC::bondlengths}, which correspond well to the cited oxidation states.  Approximately 15\% of the Tm sites were found to be randomly substituted by Mn, giving the chemical formula Tm$_{0.85}$Mn$_{3.15}$O$_6$. Five weak diffraction peaks, labeled by asterisks in Figure \ref{FIG::neutrondata}a, could not be accounted for using the $Pmmn$ crystal structure model of \ce{TmMn3O6}. Instead, they were found to originate in two impurity phases, a \ce{TmMnO3}-related phase (8.1(7) wt\%) and \ce{MnCO3} (0.5(1) wt\%). The \ce{TmMnO3} related impurity phase was determined to have composition \ce{Tm_{0.59}Mn_{0.41}MnO3}. 

\begin{table*}
{\setlength{\tabcolsep}{1.0em}
\caption{\label{TAB::xyz_bvs}Crystal structure parameters of \ce{TmMn3O6} ($Z=4$, space group $Pmmn$) refined at 85 K. The lattice parameters were determined to be $a =  7.22944(7) ~ \mathrm{\AA}$, $b = 7.41079(7)~ \mathrm{\AA}$, and $c = 7.72580(9)~ \mathrm{\AA}$. Excellent reliability parameters of $R = 3.32\%$, $wR = 2.68\%$, $R_\mathrm{Bragg} = 4.57\%$ were achieved in the refinement.}
\begin{ruledtabular}
\begin{tabular}{c | c c c c c c c}
Atom & Site & Sym. & $x$ & $y$ & $z$ & $U_\mathrm{iso}$ $(\mathrm{\AA}^2)$ & B.v.s. ($|e|$)\\
\hline
Tm1  & $2a$ & $mm2$ & 0.25 & 0.25 & 0.7844(5) & 0.0082(7) & 2.76(1)\\
Tm2  & $2a$ & $mm2$ & 0.25 & 0.25 & 0.2859(5) & 0.0082(7) & 2.74(1)\\
Mn1  & $2b$ & $mm2$ & 0.75 & 0.25 & 0.7227(7) & 0.0098(9) & 2.845(8)\\
Mn2  & $2b$ & $mm2$ & 0.75 & 0.25 & 0.2375(7) & 0.0098(9) & 2.05(1)\\
Mn3  & $4c$ & $\bar{1}$ & 0 & 0 & 0 & 0.008(1) & 3.09(2)\\
Mn4  & $4c$ & $\bar{1}$ & 0 & 0 & 0.5 & 0.006(1) & 3.59(2)\\
O1   & $8g$ & $1$   & 0.4355(2) &-0.0681(2) & 0.2671(2) & 0.0127(4) & - \\
O2   & $4f$ & $.m.$ & 0.0582(3) & 0.25 & 0.0451(3) & 0.0129(6) & - \\
O3   & $4e$ & $m..$ & 0.25 &  0.5281(3) & 0.9202(3) & 0.0087(6) & - \\
O4   & $4f$ & $.m.$ & 0.5403(3) & 0.25 & 0.4177(3) & 0.0147(7) & - \\
O5   & $4e$ & $m..$ & 0.25 & 0.4307(3) & 0.5408(3) & 0.0109(6) & - 
\end{tabular}
\end{ruledtabular}
}
\end{table*}

Below $T_\mathrm{c}$ = 74 K four new diffraction peaks appeared, which could be indexed as (011), (101), (121) and (211) with respect to the parent, $Pmmn$ crystal structure refined above, concomitant with changes in the sample's magnetisation and hence magnetic in origin. The intensities of the (002), (020) and (200) diffraction peaks, already present in the paramagnetic phase, were observed to grow below 74 K. These additional diffraction intensities were also determined to be magnetic, as their temperature dependence was the same as the first family of purely magnetic intensities listed above. All observed magnetic diffraction intensities are uniquely consistent with a \textbf{k}=(0,0,0), $\Gamma$-point magnetic propagation vector.

\begin{figure}
\includegraphics[width=\linewidth]{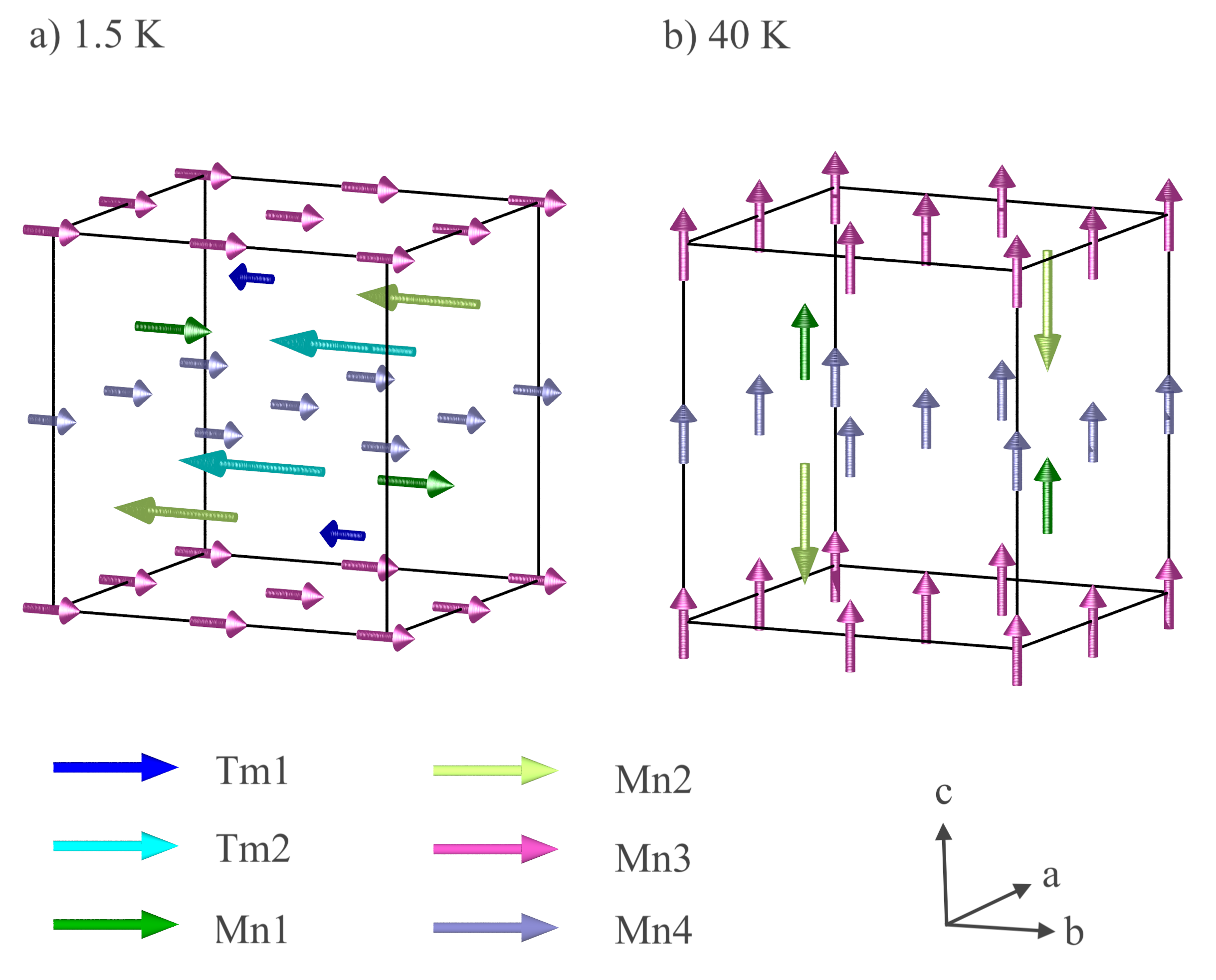}
\caption{The refined magnetic structure of \ce{TmMn3O6} at a) 1.5 K and b) 40 K. The crystallographic unit cell is drawn in black.}
\label{FIG::magstruc}
\end{figure}

The magnetic $\Gamma$-point representation for the relevant Wyckoff positions decomposes into seven irreducible representations, as given in Table \ref{TAB::irreps} of Appendix \ref{SEC::irreps} (Miller and Love notation is used throughout). Magnetic structure models constructed from linear combinations of the tabulated symmetry adapted basis functions (Table \ref{TAB::irreps}) were systematically tested against neutron powder diffraction data measured at 40 K. It was conclusively determined that between 74 and 28 K a magnetic structure involving A$'$, A$''$ and B manganese moments, which transformed by the $\Gamma_{2}^{+}$ representation, was uniquely consistent with the diffraction data. Specifically, a model in which all moments were aligned parallel to $z$ ($c$ axis), with the Mn1, Mn3 and Mn4 sublattices ferromagnetically aligned, and with the Mn2 sublattice moments antialigned (a net ferrimagnetic structure), Figure \ref{FIG::magstruc}b, had the best fit to the 40 K data shown in Figure \ref{FIG::neutrondata}b ($R = 3.90\%$, $wR = 3.09\%$, $R_\mathrm{Mag} = 4.90\%$). The refined moment magnitudes are given in Table \ref{TAB::magmoments}. The fit was found to be insensitive to interchanging the Mn1 (Mn3) and Mn2 (Mn4) moments, and we present the model most consistent with the respective oxidation states. It is noteworthy that the magnetic structure factor of the Mn B sites is exactly zero for the $(h11)$ and $(1k1)$ reflections, which therefore only have magnetic scattering contributions from the Mn A$'$ and A$''$ sites.

\begin{table}
\caption{\label{TAB::magmoments}Magnetic moments of \ce{TmMn3O6} refined at 1.5 and 40 K. Symmetry equivalent atoms have the same magnetic moment.}
\begin{ruledtabular}
\begin{tabular}{c | c c c }
Atom & 1.5 K $m||y~(\mu_\mathrm{B})$ & 40 K $m||z~(\mu_\mathrm{B})$ \\
\hline
Tm1 & -0.77(2) &  0 \\
Tm2 & -3.70(3) &  0 \\
Mn1 &  2.003(9) &  1.957(9) \\
Mn2 & -3.17(1)& -3.10(1) \\
Mn3 &  1.91(1) &  1.89(1) \\
Mn4 &  1.59(1) &  1.58(1)  
\end{tabular}
\end{ruledtabular}
\end{table}

On cooling below 28 K, the magnetic diffraction peaks undergo an approximately discontinuous change in their relative intensities. For example, the magnetic contribution to the (020) peak disappears, a large (002) magnetic intensity appears, and the (200) magnetic peak remains approximately the same. Given that the magnetic neutron diffraction intensity is proportional to the component of the spin perpendicular to the scattering vector, these changes are qualitatively consistent with a 90$^\circ$ flop of magnetic moments from the $c$ axis to the $b$ axis. Indeed, testing each of the irreducible representations against the neutron powder diffraction data conclusively revealed that the magnetic structure below $T_\mathrm{flop}=$ 28 K transformed by $\Gamma_{4}^{+}$. This single irreducible representation solution for the low temperature magnetic structure comprises the same ferrimagnetic structure found above $T_\mathrm{flop}$, but with moments aligned parallel to $y$ ($b$ axis) instead of $z$ ($c$ axis), as shown in Figure \ref{FIG::magstruc}a. In order to accurately fit the 1.5 K diffraction data shown in Figure \ref{FIG::neutrondata}c, ($R = 3.82\%$, $wR = 3.53\%$, $R_\mathrm{Mag} = 5.37\%$), it was necessary to include a finite, symmetry allowed moment on the Tm1 and Tm2 sublattices, both aligned antiparallel to the net magnetisation. The thulium contribution to the diffraction pattern is particularly apparent at $d$-spacings of 1.70-1.72{\AA}. Here the magnetic form factor of manganese is greatly reduced, and thus the observed growth in magnetic intensity below $T_\mathrm{flop}$ is principally representative of the evolution of magnetic moments on the Tm1 and Tm2 ions. The moment magnitudes for all symmetry inequivalent cations refined at 1.5 K are given in Table \ref{TAB::magmoments}.  Again, the ratios between the Mn1 (Mn3) and Mn2 (Mn4) moments were chosen to be consistent with their oxidation states (these ratios were fixed in the variable temperature data analysis presented below). We note that the Tm1 and Tm2 moments were found to be significantly smaller than their theoretical free ion values (7$\mu_\mathrm{B}$), which is consistent with a polarised singlet ground state as discussed in Section \ref{SSEC::SIA}.

As shown in Table \ref{TAB::irreps} of Appendix \ref{SEC::irreps}, both the $\Gamma_{2}^{+}$ and $\Gamma_{4}^{+}$ irreducible representations allow for a canted magnetic structure of the B sites, \emph{i.e.} the orthogonal superposition of antiferromagnetism onto the nominally ferromagnetic sublattice. However, the inclusion of these additional degrees of freedom into the refined magnetic structure model did not improve the fit, thus the collinear ferrimagnetic structures can be considered minimal models of \ce{TmMn3O6}.

Both impurity phases adopt long range magnetic order at low temperatures. Tm deficient \ce{TmMnO3} undergoes three magnetic phase transitions at $\sim$110 K, $\sim$50 K and $\sim$20 K \citep{2018Zhang}, and \ce{MnCO3} magnetically orders below 34.5 K \cite{1972Maartense}. The Tm-poor \ce{TmMnO3} magnetic structures are yet to be fully determined, so the respective intensities were fit using the Le Bail method. The \ce{MnCO3} magnetic diffraction peaks were fit using a G-type antiferromagnetic structure that well reproduced the diffraction intensities.

The temperature dependence of all symmetry inequivalent moment magnitudes was evaluated by fitting the above magnetic structure models to variable temperature neutron powder diffraction data, and is plotted in Figure \ref{FIG::tempdep}b. Critical behaviour is observed at $T_\mathrm{c}$ that is consistent with a second order phase transition.  To reliably extract the Tm moment dependences the magnitudes of the Mn moments were assumed to remain constant below 30 K. Figure \ref{FIG::tempdep}c shows the direction of the net magnetisation (moment axis) as a function of temperature, where $\Omega$ describes a rotation from the $c$ axis towards the $b$ axis. $\Omega$ was allowed to freely vary close to the spin-flop reorientation transition, but was fixed to the directions of the single irreducible representation magnetic structures deep into their respective phases.

Finally in this section, we revisit the ZFC and FCC DC $M$/$H$ measurements shown in Figure \ref{FIG::tempdep}a, and demonstrate that they are in full agreement with the results of the neutron powder diffraction data analysis. One can consider three primary contributions to the temperature dependence of $M$/$H$; i) changes in the magnetic susceptibility of the manganese ions, which is peaked at the two phase transitions \cite{2017Zhang}, ii) the Van-Vleck magnetic susceptibility of the thulium ions, which is small at all temperatures, and iii) the saturated magnetisation, domain effects, and powder averaging. Having ZFC to the lowest measured temperature both manganese and thulium susceptibilities are small, and approximately equal populations of ferrimagnetic domains cancel the sample's net magnetisation. As $T_\mathrm{flop}$ is approached on warming, the coercive field reduces 
\cite{2017Zhang} allowing for an increasing imbalance in ferrimagnetic domains induced by the small measuring field, giving rise to an increase in the sample's net magnetisation that is further enhanced by the reduction in the Tm1 and Tm2 moments aligned antiparallel to the net magnetisation. At the spin-flop transition the system becomes magnetically isotropic and the susceptibility is peaked, giving rise to a cusp in the powder averaged magnetisation. On further warming the sample maintains a net magnetisation, which decreases as $T_\mathrm{c}$ is approached, in accordance with the temperature dependence of the ordered magnetic moments (Figure \ref{FIG::tempdep}b). Above $T_\mathrm{c}$ the susceptibility follows a Curie-Weiss-like dependence of a paramagnet as show in the inset of Figure \ref{FIG::tempdep}a.

\subsection{Thulium Crystal Electric Field}\label{SSEC::CryElecField}
Tm$^{3+}$ has electronic configuration [Xe] 4$f^{12}$, and Hunds rules give S = 1, L = 5, and J = 6 for the lowest energy 13-fold multiplet of states. Tm$^{3+}$ is a non-Kramers ion, and its local crystal electric field (CEF) that transforms by the abelian point group $mm2$ (C$_{2v}$) can lift in full the multiplet degeneracy of the free ion.
The CEF Hamiltonian for a single Tm ion can be written,
\begin{equation}
\label{EQN::H_cef}
\mathcal{H}_\mathrm{cef} = \sum_{n}\sum_{m=-n}^nB_n^mO_n^m
\end{equation}
where $O_n^m$ are the Stevens operator equivalents tabulated in reference \cite{1964Hutchings}, and $B_n^m$ are the CEF parameters. For point group $mm2$, only terms with $n$ = 2, 4, and 6, and $m$ a positive even integer, are non zero by symmetry, giving 9 CEF parameters to be determined. The CEF parameters can be written as the product of three terms,
\begin{equation}
B_n^m = A_n^m\langle r^n \rangle \Theta_n
\end{equation}
where $\langle r^n \rangle$ is the expectation value of the $n^\mathrm{th}$ power of the radial distance of the Tm$^{3+}$ 4$f$ electrons, given in reference \cite{1979Freeman}, and $\Theta_n$ are the Stevens factors, given in reference \cite{1964Hutchings}. $A_n^m$ is the material specific part that describes the local CEF. Here, we estimate all 9 unknown values of $A_n^m$ by a point charge model of the form,
\begin{equation}
A_n^m = \frac{-|e|^2}{(2n+1)\epsilon_0}C_n^m\sum_i \frac{q_i}{r_i^{n+1}}Z_n^m(\theta_i,\phi_i)
\end{equation}
where the summation is taken over $i$ nearest neighbour anions, $q_i$, $r_i$, $\theta_i$, and $\phi_i$ are the charge in units $e$, and the position in spherical coordinates of the $i^\mathrm{th}$ anion, $Z_n^m$ is a tesseral harmonic, $C_n^m$ is the numerical factor occurring in $Z_n^m$, and $\epsilon_0$ is the permittivity of free space. 

The CEF parameters, $B_n^m$, were calculated for Tm1 and Tm2 using atomic fractional coordinates and lattice parameters determined at 1.5 K. The obtained values are given in Table \ref{TAB::CEF parameters} of  Appendix \ref{SEC::cef_parameters}. Similar magnitudes were obtained for the CEF parameters of the two Tm sites, but the sign of $B_2^2$, $B_4^2$, $B_6^2$, and $B_6^6$ are opposite, which can be understood based on the presence of a $4_2$ screw pseudosymmetry relating the two Tm sites. The eigenvalues and eigenfunctions of Equation \ref{EQN::H_cef} are given in Table \ref{TAB::eigen}. As expected, the Tm$^{3+}$ multiplets are split into 13 singlet states. 

\subsection{Thulium Single Ion Anisotropy}\label{SSEC::SIA}

\begin{figure}
\includegraphics[width=8.5cm]{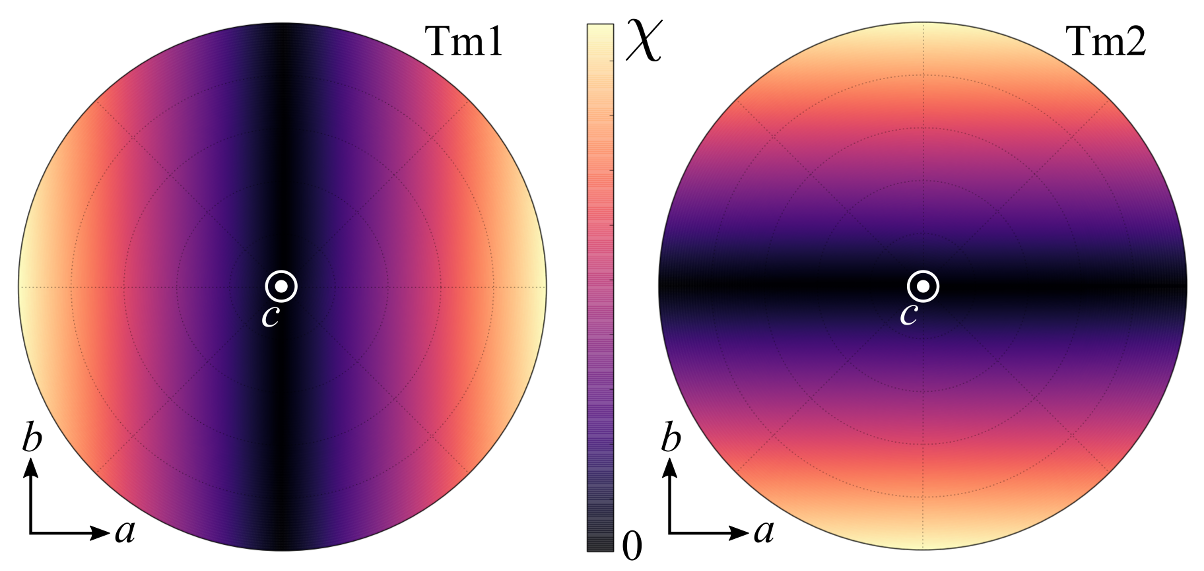}
\caption{Stereographic projections of the Tm1 and Tm2 magnetic susceptibility calculated in the limit of small applied magnetic fields. The crystal electric field induced Ising-like single ion anisotropy of Tm1 and Tm2 lies parallel to $a$ and $b$, respectively.}
\label{FIG::SIA}
\end{figure}

The 0 K singlet ground state is non-magnetic. However, there does exist an instability towards a polarised magnetic state when under a magnetic field, $\mathbf{B}$ \cite{1971Cooper}. The polarising field may originate in $f$-$d$ magnetic exchange interactions between Tm and the Mn sublattice, the dipolar field of the Mn sublattice, or an externally applied magnetic field. The dipolar field of Mn can be ruled out, as it gives moment directions inconsistent with the experimentally determined magnetic structure. We therefore extend the single ion Hamiltonian;
\begin{equation}\label{EQN::H_single}
\mathcal{H} = \mathcal{H}_\mathrm{cef} + g_j\mu_\mathrm{B}\mathbf{J}\cdot\mathbf{B}_{fd}
\end{equation}
where $\mathbf{B}_{fd}$ represents an effective $f$-$d$ exchange field. The sum of expectation values, $\langle$J$_x\rangle$, $\langle$J$_y\rangle$, and $\langle$J$_z\rangle$, for all eigenstates, weighted by the Boltzmann distribution ($T=10$ K), was evaluated for numerous magnetic field directions covering a full hemisphere, in the limit of small $\mathbf{B}_{fd}$. A strong Ising-like single ion anisotropy (SIA) was found for both Tm ions, with the Ising directions of Tm1 and Tm2 parallel to the crystallographic $a$ and $b$ axes, respectively, as illustrated by the stereographic projections shown in Figure \ref{FIG::SIA}. We note that the 90$^\circ$ rotation of the Ising axis from Tm1 to Tm2 is consistent with the tetragonal $4_2$ screw pseudo-symmetry discussed above.

The magnetic moments of the Mn B site sublattice, which provides the uniaxial polarising field for both Tm ions via $f$-$d$ exchange, were found to be aligned parallel to the $b$ axis in the ground state. In this case one would expect a large moment to be induced on Tm2, and a very small moment to be induced on Tm1. The former is fully consistent with our neutron powder diffraction results. However, the sizeable magnetic moment of Tm1 found perpendicular to its Ising axis cannot be explained by $f$-$d$ exchange alone, at least from the results of our calculations. We suggest that either the larger Tm1 moment could naturally occur due to finite, ferromagnetic $f$-$f$ exchange interactions, or because the magnitude of the SIA has been overestimated. 

Finally in this section we note that the point charge model used above does not fully describe the true crystal electric field as it omits orbital considerations. However, one can show that such a model is sufficient to robustly determine the Tm single ion anisotropy in \ce{TmMn3O6}. The sign of $B_2^0$ determines the single ion anisotropy to be Ising-like parallel to $z$ (negative), or easy plane perpendicular to $z$ (positive). In the latter case, large $B_2^2$ then imposes Ising-like anisotropy in the $xy$ plane, the direction of which is determined by the sign of $B_2^2$ (negative makes it parallel to $x$ and positive parallel to $y$). Given the dominance of these terms (Table \ref{TAB::CEF parameters}) in our approximation of the CEF, any perturbation away from this model will not affect the qualitative determination of the Ising-like SIA and its direction. 

\section{Discussion}\label{SEC::discussion}
First, we consider the role of B-B exchange interactions. The Jahn-Teller active, octahedral oxygen coordinations of the B site Mn$^{3+}$ ions (Mn3) are elongated along $z$, lifting the degeneracy of the $e_{g}$ states and giving preferential electronic occupation of the $d_{3z^2-r^2}$ orbitals --- a ferro-orbital order. Thus for Mn3 ions, all orbitals are half occupied except for the $d_{x^2-y^2}$ orbital which is empty. In comparison, the B site Mn$^{3.5+}$ octahedra (Mn4) do not have a strong Jahn-Teller distortion, and can be considered to have singly occupied $t_{2g}$ orbitals and a fractional electronic occupation of the $e_{g}$ orbitals. According to the Goodenough-Kanamori-Anderson (GKA) rules, if exchange is mediated between 180{\degree} cation-anion-cation bonds that have $p\sigma$ and $p\pi$ superexchange, which can be considered an approximation of the bonds \emph{in} the Mn3 and Mn4 layers, then one expects the exchange interactions to be antiferromagnetic. The GKA rules further predict that the exchange interaction \emph{between} Mn3 and Mn4 layers, which is mediated by $\sim$180{\degree} cation-anion-cation interactions between half-occupied and empty $e_g$ states through $p\sigma$ superexchange, will be ferromagnetic. The result is a so-called C-type antiferromagnetic structure, in contradiction with the experimentally determined ferromagnetic B site sublattice.

As a direct consequence of the $mm2$ point symmetry of the A, A$'$ and A$''$ sites, all four nearest neighbour A-B exchange interactions must be equivalent (see Figure \ref{FIG::crystruc}e), and hence can only lower the energy of the system if ferromagnetism resides in the B site layers. To the contrary, the antiferromagnetic order favoured by B-B exchange exactly frustrates the A-B exchange. Given the experimentally determined ferrimagnetic structure, it is apparent that in \ce{TmMn3O6} A-B exchange dominates B-B exchange  --- an unexpected scenario that is emerging in the study of quadruple perovskites \cite{2018Johnson_1}, where large deviations from 180{\degree} B-O-B bond angles significantly weaken the respective superexchange interactions (in \ce{TmMn3O6} these bond angles are $\sim$140{\degree}, as given in Table \ref{TAB::Bsitebonds} of Appendix \ref{SEC::bondlengths}). One might expect that the \ce{$R$Mn3O6} system could be tuned from ferrimagnetic to antiferromagnetic by modifying the critical balance between A-B and B-B exchange. In the absence of significant A-A exchange the stability of these two magnetic structures does not, in principle, depend upon long range magnetic order of the A, A$'$ or A$''$ ions. Hence this tuning can be achieved at the local level, where individual spins contribute on average to a net A-B exchange interaction. 

We now consider the microscopic origin of the spin-flop transition. Above $T_\mathrm{flop}$ the direction of the magnetic moments is determined by the average Mn single ion anisotropy, which is likely dominated by the electronic configuration of the Jahn-Teller active Mn3 ions placing the moments along $c$ \cite{2015Whangbo}. However, the $c$ axis is a magnetically hard axis for both Tm1 and Tm2, whose Ising-like SIA was calculated to be along the $a$ and $b$ axes, respectively (Section \ref{SSEC::SIA}). Hence, as the system temperature reaches the energy scale of the $f$-$d$ exchange field \cite{2009Pomjakushin, 2009Salama}, it becomes energetically favourable for the manganese magnetic moments to flop into the $ab$-plane such that the Tm ions adopt a polarised magnetic moment (at the expense of the weaker manganese SIA). The magnitudes of the polarised Tm moments, and the global in-plane magnetisation direction, are then respectively determined by the absolute and relative magnitude of the $f$-$d$ exchange field at the Tm1 and Tm2 sites. In \ce{TmMn3O6} the Tm2 moment dominates below $T_\mathrm{flop}$, imposing its $b$ axis SIA on the system. Hence we conclude that the $f$-$d$ exchange fields in \ce{TmMn3O6} are significantly larger at the Tm2 site than at the Tm1 site. In this regard, it is interesting to consider the $4_2$ screw pseudo symmetry discussed in the theoretical sections above. If the $4_2$ screw is present, the Tm sites still have $mm2$ point symmetry, but their crystal electric field and single ion anisotropy become equivalent, related by a 90$^\circ$ rotation in the $ab$ plane. In this case one can show that a spin-flop transition will still occur, but the Tm sites adopt orthogonally polarised moments with Mn moments bisecting the two. In \ce{TmMn3O6} the $4_2$ screw symmetry is broken by charge and orbital order \cite{2017Zhang}, and it is exactly this charge and orbital order that gives rise to significantly different nearest neighbour Tm-O-Mn bonding conditions for Tm1 and Tm2, resulting in different $f$-$d$ exchange fields at the two sites, and hence allowing a collinear ground state ferrimagnetic structure.

\section{Conclusions}\label{SEC::conclusions}
In \ce{TmMn3O6}, all A$'$, A$''$ and B site manganese ions develop a collinear ferrimagnetic structure below 74 K, with moments orientated along the $c$ axis. Remarkably, the ferrimagnetic order is in contradiciton to that predicted by Goodenough-Kanamori-Anderson rules, whereby ferro-orbital order favours C-type antiferromagnetism instead. In \ce{TmMn3O6}, the average A-B exchange, together with the $mm2$ A, A$'$ and A$''$ site symmetry, wins over B-B exchange and gives rise to ferrimagnetism. This offers the possibility of tuning the magnetic structure of \ce{$R$Mn3O6} oxides from ferrimagnetic to antiferromagnetic through a modification of the competition between A-B and B-B exchange, for instance through doping of the A, A$'$ or A$''$ sites. Below 28 K, the Mn moments flop through a spin reorientation transition, retaining the higher temperature ferrimagnetic structure but with moments aligned along the $b$ axis, and with a finite moment developing on both Tm sublattices aligned antiparallel to the net Mn magnetisation. Point charge calculations show that the Tm1 and Tm2 sublattices can adopt a polarised magnetic moment lying within the $ab$-plane, induced by a uniaxial $f$-$d$ exchange field created by the ordered Mn B site sublattice, which motivates the spin flop transition at low temperature. The magnetic behaviour of \ce{TmMn3O6} is strikingly similar to that of the rare-earth orthoferrites (\ce{$R$FeO3}), which crystallise in the orthorhombic $Pbnm$ space group, and feature spin reorientation transitions driven by the single ion anisotropy of the rare earth ions. It is therefore not far to imagine that the novel \ce{$R$Mn3O6} compounds, like the rare-earth orthoferrites, may feature laser \cite{2004Kimel}, temperature \cite{2014Cao} and or applied magnetic field \cite{1980Johnson}, spin-reorientation transitions for applications in spintronics, and also possibly magnonics \cite{2018Grishunin}.

\section{Acknowledgements}
R. D. J. acknowledges financial support from the Royal Society. P. G. R. and R. D. J. acknowledge support from
EPSRC, Grant No. EP/M020517/1, entitled “Oxford Quantum Materials Platform Grant.” This study was supported in part by JSPS KAKENHI Grant Number JP16H04501, and a research grant from Nippon Sheet Glass Foundation for Materials and Engineering, and Innovative Science and Technology Initiative for Security, ATLA, Japan. 

\bibliography{tmopaper}

\appendix
\section{\label{SEC::bondlengths}Selected bond lengths and angles for \ce{TmMn3O6}}
We give selected cation-anion bond lengths for the A, A$'$, A$''$ cations in Table \ref{TAB::Asitebonds} and selected cation-anion bond lengths and cation-anion-cation bond angles for the B site cations in Table \ref{TAB::Bsitebonds}.

\begin{table*}
{\setlength{\tabcolsep}{1.0em}
\caption{\label{TAB::Asitebonds} Bond lengths for the A, A$'$ and A$''$ sites refined at 85K, for \ce{TmMn3O6} in the $Pmmn$ space group.}
\begin{ruledtabular}
\begin{tabular}{c c | c c | c c }
Tm1-O1 ($\times 4$) & 2.672(2) {\AA}  &  Tm2-O1 ($\times 4$) & 2.716(2) {\AA}   & Mn1-O1 ($\times 4$) & 1.903(2) {\AA} \\ 
Tm1-O2 ($\times 2$) & 2.445(4) {\AA}  &  Tm2-O2 ($\times 2$) & 2.320(4) {\AA}   &                     &                \\ 
Tm1-O3 ($\times 2$) & 2.313(3) {\AA}  &  Tm2-O4 ($\times 2$) & 2.333(3) {\AA}   & Mn2-O2 ($\times 2$) & 2.678(4) {\AA} \\ 
Tm1-O5 ($\times 2$) & 2.310(4) {\AA}  &  Tm2-O5 ($\times 2$) & 2.381(4) {\AA}   & Mn2-O4 ($\times 2$) & 2.058(5) {\AA} \\  
\end{tabular}
\end{ruledtabular}
}
\end{table*}

\begin{table*}
{\setlength{\tabcolsep}{1.0em}
\caption{\label{TAB::Bsitebonds} Bond lengths and selected bond angles for the B-sites refined at 85K, for \ce{TmMn3O6} in the $Pmmn$ space group.}
\begin{ruledtabular}
\begin{tabular}{c c | c c | c c }
Mn3-O1 ($\times 2$) & 2.1748(2) {\AA} & Mn4-O1 ($\times 2$)& 1.9259(2) {\AA} & Mn3-O2-Mn3 ($\times 2$) & 147.1(1){\degree}\\ 
Mn3-O2 ($\times 2$) & 1.9316(7) {\AA} & Mn4-O4 ($\times 2$)& 1.9803(9) {\AA} & Mn3-O3-Mn3 ($\times 2$) & 140.4(1){\degree}\\ 
Mn3-O3 ($\times 2$) & 1.9209(8) {\AA} & Mn4-O5 ($\times 2$)& 1.9052(8) {\AA} & Mn3-O1-Mn4 ($\times 2$) & 140.7(8){\degree}\\ 
				    & 				  & 				   & 				 & Mn4-O4-Mn4 ($\times 2$) & 138.6(1){\degree}\\
				    & 				  &				       &				 & Mn4-O5-Mn4 ($\times 2$) & 143.1(1){\degree}\\				
\end{tabular}
\end{ruledtabular}
}
\end{table*}

\section{\label{SEC::irreps}Irreducible representations and their symmetry adapted basis functions}
We give the basis functions of the $\Gamma$-point irreducible representations used to describe the \ce{TmMn3O6} magnetic structures (Table \ref{TAB::irreps}). The symmetry analysis was performed using the \textsc{isotropy} suite \cite{2006Isodistort}.

\begin{table*}[!h]
\caption{\label{TAB::irreps}Basis functions of the seven, one dimensional $\Gamma$-point irreducible representations that appear in the decomposition of the full magnetic representation for the relevant Wyckoff positions of \ce{TmMn3O6}. The $+$ and $-$ symbols denote the relative sign of magnetic moment components, $x$, $y$, $z$, on every sublattice of symmetry equivalent atoms. The symbols in bold highlight the magnetic moment components used to fit the magnetic structures in section \ref{SEC::results}, $\Gamma_{2}^{+}$ for the first magnetic phase and $\Gamma_{4}^{+}$ for the second magnetic phase ($T \textless T_\mathrm{flop}$). The moment components of the symmetry equivalent atoms have the same magnitude. The relative signs and magnitudes of orthogonal components, or of moments on inequivalent atomic sublattices, are not determined by symmetry. The absence of a $+$ or $-$ sign indicates that that component is zero by symmetry.}
\begin{ruledtabular}
\begin{tabular}{c | c | c c c | c c c | c c c | c c c | c c c | c c c | c c c}
Atom  & Frac. coords. &  \multicolumn{3}{c}{$\Gamma_{1}^{+}$} & \multicolumn{3}{c}{$\Gamma_{2}^{+}$} &\multicolumn{3}{c}{$\Gamma_{3}^{+}$} & \multicolumn{3}{c}{$\Gamma_{4}^{+}$}& \multicolumn{3}{c}{$\Gamma_{1}^{-}$} & \multicolumn{3}{c}{$\Gamma_{3}^{-}$}& \multicolumn{3}{c}{$\Gamma_{4}^{-}$} \\
& & $x$ & $y$ & $z$ & $x$ & $y$ & $z$ & $x$ & $y$ & $z$ & $x$ & $y$ & $z$ & $x$ & $y$ & $z$ & $x$ & $y$ & $z$ & $x$ & $y$ & $z$ \\
\hline
Tm1 & 0.25, 0.25, $ z$ & & & & & & \textbf{+} & + & & & & \textbf{+} & & & & + & & + & & + & & \\
    & 0.75, 0.75, $-z$ & & & & & & \textbf{+} & + & & & & \textbf{+} & & & & - & & - & & - & & \\
	& & & & & & & & & & & & & & & & & & & & & &\\
Tm2 & 0.25, 0.25, $ z$ & & & & & & \textbf{+} & + & & & & \textbf{+} & & & & + & & + & & + & & \\
    & 0.75, 0.75, $-z$ & & & & & & \textbf{+} & + & & & & \textbf{+} & & & & - & & - & & - & & \\
& & & & & & & & & & & & & & & & & & & & & &\\
Mn1 & 0.75, 0.25, $ z$ & & & & & & \textbf{+} & + & & & & \textbf{+} & & & & + & & + & & + & & \\
    & 0.25, 0.75, $-z$ & & & & & & \textbf{+} & + & & & & \textbf{+} & & & & - & & - & & - & & \\
& & & & & & & & & & & & & & & & & & & & & &\\
Mn2 & 0.75, 0.25, $ z$ & & & & & & \textbf{+} & + & & & & \textbf{+} & & & & + & & + & & + & & \\
    & 0.25, 0.75, $-z$ & & & & & & \textbf{+} & + & & & & \textbf{+} & & & & - & & - & & - & & \\
& & & & & & & & & & & & & & & & & & & & & &\\
Mn3 & 0.5, 0, 0 & + & + & + & + & + & \textbf{+} & + & + & + & + & \textbf{+} & + & & & & & & & & & \\
    & 0, 0.5, 0 & - & - & + & - & - & \textbf{+} & + & + & - & + & \textbf{+} & - & & & & & & & & & \\
    & 0, 0, 0 & + & - & - & - & + & \textbf{+} & + & - & - & - & \textbf{+} & + & & & & & & & & & \\
    & 0.5, 0.5, 0 & - & + & - & + & - & \textbf{+} & + & - & + & - & \textbf{+} & - & & & & & & & & & \\
& & & & & & & & & & & & & & & & & & & & & &\\
Mn4 & 0, 0.5, 0.5 & + & + & + & + & + & \textbf{+} & + & + & + & + & \textbf{+} & + & & & & & & & & & \\
         & 0.5, 0, 0.5 & - & - & + & - & - & \textbf{+} & + & + & - & + & \textbf{+} & - & & & & & & & & & \\
         & 0.5, 0.5, 0.5 & + & - & - & - & + & \textbf{+} & + & - & - & - & \textbf{+} & + & & & & & & & & & \\
         & 0, 0, 0.5 & - & + & - & + & - & \textbf{+} & + & - & + & - & \textbf{+} & - & & & & & & & & & \\
\end{tabular}
\end{ruledtabular}
\end{table*}

\section{\label{SEC::cef_parameters}Thulium crystal electric field parameters, eigenvalues, and eigenstates}

We give the crystal electric field parameters for Tm1 and Tm2, calculated using the point charge model described in the main text (Table \ref{TAB::CEF parameters}), and the eigenstates and respective energy eignevalues for the 13-fold manifold of states (Table \ref{TAB::eigen}).

\begin{table*}[!h]
\caption{\label{TAB::CEF parameters}Crystal electric field parameters for Tm1 and Tm2, evaluated by the point charge model, and given in units $\mu$eV.}
\begin{ruledtabular}
\begin{tabular}{c | c c c c c c c c c}
Atom & $B_2^0$ & $B_2^2$ & $B_4^0$ & $B_4^2$ & $B_4^4$ & $B_6^0$ & $B_6^2$ & $B_6^4$ & $B_6^6$\\
\hline
Tm1 & 1250 &  -915 & -0.445 & -3.70 & 2.53 & 0.00299 & -0.0185 & -0.0522 &  0.0495 \\
Tm2 & 1252 &  741 & -0.391 &  3.12 & 3.12 & 0.00353 &  0.0254 & -0.0521 & -0.0468 \\
\end{tabular}
\end{ruledtabular}
\end{table*}

\begin{table*}[!h]
\caption{\label{TAB::eigen}Energy and eigenfunctions of the 13 singlet states of both Tm ions, as calculated using the crystal electric field Hamiltonian.}
\begin{ruledtabular}
\begin{tabular}{c | c c }
Atom & Energy (meV) & Eigenfunction\\
\hline
Tm1\\
&   0    &  0.005 $|6,\pm6\rangle$ $+$ 0.056 $|6,\pm4\rangle$ $+$ 0.390 $|6,\pm2\rangle$ + 0.831 $|6,0\rangle$ \\ 
&   0.20 & 0.018 $|6,\pm5\rangle$ $+$ 0.167 $|6,\pm3\rangle$ $+$ 0.687 $|6,\pm1\rangle$\\
&  22.90 & $\mp$0.076 $|6,\pm5\rangle$ $\mp$ 0.405 $|6,\pm3\rangle$ $\mp$ 0.574 $|6,\pm1\rangle$ \\
&  24.52 & $\mp$0.027 $|6,\pm6\rangle$ $\mp$ 0.211 $|6,\pm4\rangle$ $\mp$ 0.674 $|6,\pm2\rangle$ \\
&  46.79 & $-$0.040 $|6,\pm6\rangle$ $-$ 0.262 $|6,\pm4\rangle$ $-$ 0.535  $|6,\pm2\rangle$ $+$ 0.537 $|6,0\rangle$ \\
&  55.42 & 0.176 $|6,\pm5\rangle$ $+$ 0.665 $|6,\pm3\rangle$ $-$ 0.166 $|6,\pm1\rangle$ \\
&  63.20 & $\mp$ 0.156 $|6,\pm5\rangle$ $\mp$ 0.554 $|6,\pm3\rangle$ $\pm$ 0.411 $|6,\pm1\rangle$ \\
 &  84.42 & $\pm$0.118 $|6,\pm6\rangle$ $\pm$ 0.664 $|6,\pm4\rangle$ $\mp$ 0.212 $|6,\pm2\rangle$ \\
 &  84.88 & 0.113 $|6,\pm6\rangle$ $+$ 0.643 $|6,\pm4\rangle$ $-$ 0.250 $|6,\pm2\rangle$ $+$ 0.147 $|6,0\rangle$ \\
& 116.24 & $\pm$ 0.685 $|6,\pm5\rangle$ $\mp$ 0.171 $|6,\pm3\rangle$ $\pm$ 0.030 $|6,\pm1\rangle$ \\
& 116.40 & 0.685 $|6,\pm5\rangle$ $-$ 0.175 $|6,\pm3\rangle$ $+$ 0.024 $|6,\pm1\rangle$ \\
& 153.17 & 0.697 $|6,\pm6\rangle$ $-$ 0.120 $|6,\pm4\rangle$ $+$ 0.007 $|6,\pm2\rangle$ $+$ 0.001 $|6,0\rangle$ \\
& 153.18 & $\mp$0.697 $|6,\pm6\rangle$ $\pm$ 0.121 $|6,\pm4\rangle$ $\mp$ 0.010 $|6,\pm2\rangle$ \\ 
\\
\hline
Tm2\\
&  0     & $-$0.002 $|6,\pm6\rangle$ $+$ 0.032 $|6,\pm4\rangle$ $-$ 0.357 $|6,\pm2\rangle$ $+$ 0.862 $|6,0\rangle$  \\
&   0.32 & $\pm$0.009 $|6,\pm5\rangle$ $\mp$ 0.130 $|6,\pm3\rangle$ $\pm$ 0.695 $|6,\pm1\rangle$  \\
&  19.49 & 0.057 $|6,\pm5\rangle$ $-$ 0.363 $|6,\pm3\rangle$ $+$ 0.604 $|6,\pm1\rangle$   \\
&  21.65 & $\pm$ 0.021 $|6,\pm6\rangle$ $\mp$ 0.176 $|6,\pm4\rangle$ $\pm$ 0.685 $|6,\pm2\rangle$  \\
&  40.91 & 0.029 $|6,\pm6\rangle$ $-$ 0.215 $|6,\pm4\rangle$ $+$ 0.576 $|6,\pm2\rangle$ $+$ 0.493 $|6,0\rangle$  \\
&  50.91 & $\pm$0.146 $|6,\pm5\rangle$ $\mp$ 0.680 $|6,\pm3\rangle$ $\mp$ 0.129 $|6,\pm1\rangle$  \\
&  56.45 & $-$0.131 $|6,\pm5\rangle$ $+$ 0.590 $|6,\pm3\rangle$ $+$ 0.367 $|6,\pm1\rangle$  \\
&  78.69 & $\mp$ 0.100 $|6,\pm6\rangle$ $\pm$ 0.677 $|6,\pm4\rangle$ $\pm$ 0.177 $|6,\pm2\rangle$ \\
&  78.84 & $-$0.097 $|6,\pm6\rangle$ $+$ 0.665 $|6,\pm4\rangle$ $+$ 0.203 $|6,\pm2\rangle$ $+$ 0.118 $|6,0\rangle$  \\
& 110.48 & 0.693 $|6,\pm5\rangle$ $+$ 0.141 $|6,\pm3\rangle$ $+$ 0.020 $|6,\pm1\rangle$   \\
& 110.60 & $\mp$0.692 $|6,\pm5\rangle$ $\mp$ 0.145 $|6,\pm3\rangle$ $\mp$ 0.018 $|6,\pm1\rangle$  \\
& 149.07 & 0.700 $|6,\pm6\rangle$ $+$ 0.102 $|6,\pm4\rangle$ $+$ 0.003 $|6,\pm2\rangle$ $-$ 0.002 $|6,0\rangle$ \\
& 149.08 & $\pm$0.700 $|6,\pm6\rangle$ $\pm$ 0.102 $|6,\pm4\rangle$ $\pm$ 0.005 $|6,\pm2\rangle$
\end{tabular}
\end{ruledtabular}
\end{table*}

\end{document}